\documentclass[conference]{IEEEtran}
\IEEEoverridecommandlockouts
\usepackage[utf8]{inputenc}
\usepackage{cite}
\usepackage{amsmath,amssymb,amsfonts}
\usepackage{algpseudocode}%
\usepackage{graphicx}
\usepackage{textcomp}
\usepackage{algorithm}
\usepackage{url}
\usepackage{xcolor}
\def\BibTeX{{\rm B\kern-.05em{\sc i\kern-.025em b}\kern-.08em
    T\kern-.1667em\lower.7ex\hbox{E}\kern-.125emX}}
\begin{document}

\title{Opt-GPTQ: An Optimized GPTQ Combining Sparse Attention and Quantization Techniques}

\author{ 
\IEEEauthorblockN{1\textsuperscript{st} Jie Kong}
\IEEEauthorblockA{
\textit{School of Computer Science} \\
\textit{and Engineering} \\
\textit{Shandong University of}\\
\textit{Science and Technology}\\
Qingdao, China \\
jiekong0112@gmail.com}\and

\IEEEauthorblockN{2\textsuperscript{nd} Junxiang Zhang}
\IEEEauthorblockA{
\textit{School of Computer Science} \\
\textit{and Engineering} \\
\textit{Shandong University of}\\
\textit{Science and Technology}\\
Qingdao, China \\
junxiangzhang@sdust.edu.cn}
\and

\IEEEauthorblockN{3\textsuperscript{rd} Jiheng Xu}
\IEEEauthorblockA{
\textit{School of Computer Science} \\
\textit{and Engineering} \\
\textit{Shandong University of}\\
\textit{Science and Technology}\\
Qingdao, China \\
jihengxu@sdust.edu.cn}
\and

\IEEEauthorblockN{4\textsuperscript{th} Yalong Li}
\IEEEauthorblockA{
\textit{School of Computer Science} \\
\textit{and Engineering} \\
\textit{Shandong University of Science}\\
\textit{and Technology}\\
Qingdao, China \\
yalongli@sdust.edu.cn}
\and

\IEEEauthorblockN{5\textsuperscript{th} Shouhua Zhang}
\IEEEauthorblockA{
\textit{Faculty of Information} \\
\textit{Technology and Electrical} \\
\textit{Engineering} \\
\textit{University of Oulu} \\
Oulu, Finland \\
shouhua.zhang@oulu.fi}
\and

\IEEEauthorblockN{6\textsuperscript{th} Jiehan Zhou\textsuperscript{*}}
\IEEEauthorblockA{
\textit{School of Computer Science} \\
\textit{and Engineering} \\
\textit{Shandong University of Science}\\
\textit{and Technology}\\
Qingdao, China \\
jiehan.zhou@sdust.edu.cn}
\and

\IEEEauthorblockN{7\textsuperscript{th} Yuhai Liu}
\IEEEauthorblockA{
\textit{Dawning Information Industry} \\
\textit{Co., Ltd} \\
Qingdao, China \\
liuyh1@sugon.com}
\and

\IEEEauthorblockN{8\textsuperscript{th} Peng Liang}
\IEEEauthorblockA{
\textit{School of Chemistry and} \\
\textit{Bioengineering} \\
\textit{Shandong University of Science}\\
\textit{and Technology}\\
Qingdao, China \\
liangpeng202@hotmail.com}
\and

\IEEEauthorblockN{9\textsuperscript{th} Quan Zhang}
\IEEEauthorblockA{
\textit{School of Computer Science and} \\
\textit{Engineering} \\
\textit{Southwest Petroleum University}\\
Chengdu, China \\
zhangquan@swpu.edu.cn}
\and

\IEEEauthorblockN{10\textsuperscript{th} Luohan Jiang}
\IEEEauthorblockA{
\textit{school of Computer Science and} \\
\textit{Engineering} \\
\textit{Shandong University of Science}\\
\textit{and Technology}\\
Qingdao, China \\
luo.hj@foxmail.com}
}

\maketitle

\begin{abstract}
In the field of deep learning, traditional attention mechanisms face significant challenges related to high computational complexity and large memory consumption when processing long sequence data. To address these limitations, we propose Opt-GPTQ, an optimized Gradient-based Post Training Quantization (GPTQ) combining the Grouped Query Attention (GQA) mechanism with paging memory management, optimizing the traditional Multi-Head Attention (MHA) mechanism by grouping query heads and sharing key-value vectors. Optimized GQA (Opt-GQA) effectively reduces computational complexity, minimizes memory fragmentation, and enhances memory utilization for large-scale models. Opt-GPTQ is optimized for Data Center Units (DCUs) and integrated into the vLLM model to maximize hardware efficiency. It customizes GPU kernels to further enhance attention computation by reducing memory access latency and boosting parallel computing capabilities. Opt-GQA integrates Attention with Linear Biases (ALiBi) to reduce overhead and enhance long-sequence processing. Experimental results show that Opt-GPTQ significantly reduces computation time and memory usage while improving model performance.
\end{abstract}


\section{Introduction}
In recent years, the increasing application of artificial intelligence (AI) has driven the development and deployment of large-scale deep learning models, particularly large language models (LLMs) \cite{b1,b2}. These models, which contain billions of parameters, exhibit exceptionally high computational complexity, necessitating specialized hardware for efficient training and inference.
\par
As an emerging high-performance computing platform, DCUs have garnered significant attention in China, owing to their cost-effectiveness and computational resource advantages. However, research focused on optimizing DCUs is still in its infancy, though some exploratory work has already attracted interest \cite{b3,b4}. These studies suggest that DCUs show considerable potential for high-performance computing and the optimization of large-scale scientific models. Nevertheless, research specifically targeting the optimization of LLMs on DCUs, particularly concerning attention mechanisms, remains limited.
\par
The MHA mechanism is a core component of LLMs, used to capture complex dependencies in data. However, recent studies have highlighted several shortcomings of MHA: inefficiencies in resource usage and insufficient expressive power, especially when each attention head processes input data independently, leading to redundant computations across different heads \cite{b5,b6}; limited capacity for modeling positional information in relation to the queries and keys, which hampers the ability to capture long-range dependencies \cite{b7}; and high computational complexity, resulting in significant resource consumption, particularly during the training and inference phases of large language models. Consequently, restructuring the attention mechanism to optimize the operation and performance of large language models on DCUs has become a critical research topic in both computer science and applied computational fields.
\par
To address the issues of resource efficiency and computational complexity in the multi-head attention mechanism, GQA proposes an optimization scheme based on fixed grouping \cite{b8}. GQA divides all query heads into several predefined groups, with each group sharing a set of key and value vectors, thereby reducing redundant key-value pair computations and storage overhead \cite{b9}. Query heads within the same group perform attention computations collectively, effectively lowering overall computational complexity and improving inference efficiency. However, due to the static and fixed grouping strategy, GQA still has limitations in terms of expressive flexibility and model generalization capabilities, particularly when dealing with dynamic input features or distributed hardware architectures (such as DCUs), where there is room for improvement in resource scheduling and performance utilization.
\par
Therefore, to further improve GQA to adapt to DCU architectures and enhance its computational efficiency and expressive capabilities in actual deployment, we propose the Opt-GQA mechanism.
\par
Our contributions are as follows.
\begin{itemize}
\item We propose the Opt-GQA to replace the traditional
MHA. Opt-GQA reduces redundancy by grouping
queries and interacting with key-value pairs in parallel,
improving efficiency and reducing resource
consumption. Its parallel nature enhances
compatibility with DCUs.
\item We implemented Opt-GQA on DCU within vLLM, improving throughput and memory efficiency. Only slight latency increases were observed in a few models, while fully utilizing the capabilities of the DCU and preserving accuracy.
\end{itemize}

\section{PRINCIPLES AND MECHANISMS}
In this section, we provide a comprehensive introduction to the design  of Opt-GQA. which is an optimization method for traditional multi-head attention mechanisms in high-concurrency inference scenarios, as in Fig. 1.

\begin{figure}[htbp]
	\begin{center}
		\includegraphics[width=\linewidth]{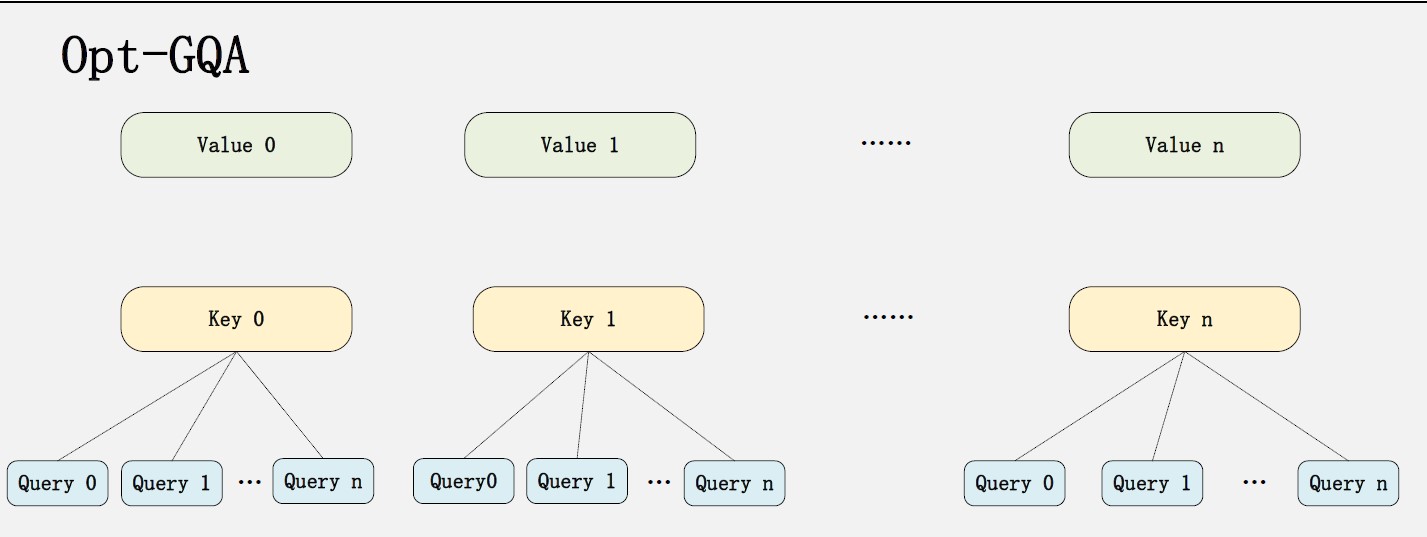}
	\end{center}
	\caption{Opt-GQA schematic diagram}
	\label{fig}
\end{figure}

In traditional MHA, the input query, key, and value
vectors are divided into multiple attention heads, with each
head performing attention calculations independently. While
this independence enhances the model's representational
capacity, it also incurs high computational and memory
overhead, especially when dealing with large-scale models
and high-concurrency inference. Opt-GQA optimizes MHA in
the following ways:
\par

\begin{itemize}

\item \textbf{Query Grouping}
\par
We first determine an appropriate grouping strategy based on the hardware architecture and resource characteristics of the DCU. For DCUs with higher computational power and memory bandwidth, we employ a larger number of groups to enhance parallelism; whereas on resource-constrained DCUs, we use fewer groups to optimize memory utilization. Specifically, the grouping strategy depends not only on the number of groups but also on the number of query heads per group: when parallel computing units are abundant, the number of heads per group can be increased to improve utilization; conversely, in memory-bandwidth-limited scenarios, the number of heads per group should be reduced to avoid bandwidth bottlenecks.

\item \textbf{Shared Key-Value}
\par
To enable the sharing of key and value vectors, we
introduce a dedicated cache management mechanism in
the vLLM model. Specifically, The G queries within each group share a set of key and value vectors. This means that multiple 
heads within the same group do not need to independently store and compute  their  own  key-value  
vectors, This approach ensures that the shared vectors are managed effectively while optimizing memory usage for large-scale models, reducing memory usage and computational redundancy. \cite{b10}.

\item \textbf{Integration and Optimization of ALiBi}
\par
Furthermore, vLLM integrates the ALiBi (Attention with Linear Biases) mechanism to improve the efficiency of attention computation. ALiBi introduces a linear bias based on the relative positions between query-key pairs, effectively replacing conventional causal masking. This approach eliminates the need to construct large mask matrices, thereby reducing both memory usage and computational overhead. The bias is directly added to the attention scores, enabling more efficient computation without explicit masking. In the parallel execution of Opt-GQA, ALiBi further enhances the efficiency of query heads accessing shared key-value pairs, reducing memory access latency. By combining custom DCU kernels with ALiBi’s positional bias, vLLM achieves higher throughput and lower memory overhead, particularly in long-sequence GQA scenarios.

\item \textbf{Efficient Parallel Computation}
\par
By reducing the redundant storage and computation of
key-value vectors, Opt-GQA can make more efficient
use of hardware resources, especially on parallel
computing platforms such as DCUs, thereby
improving overall computational throughput.
\end{itemize}

Opt-GQA optimizes traditional GQA by grouping query vectors and sharing key-value vectors within each group. This reduces computational redundancy and memory usage, making it more efficient \cite{b11}, especially in high-concurrency inference scenarios. By leveraging dynamic grouping based on activation similarity, Opt-GQA improves computational throughput, scalability, and model performance \cite{b12, b13}. Compared to traditional MHA, and this new method enhances efficiency, memory utilization, and parallel processing, supporting more efficient inference for large-scale models.

\section{METHOD}
To efficiently run LLMs on DCUs, we integrate the Opt-GQA mechanism into the vLLM model by using the DTK library, The specific implementation steps are as follows:

\subsection{Opt-GQA Reasoning Process}

First, the input Query, Key and Value tensor is reshaped to fit the shape of the attention computation. When processing grouped queries, we introduce the paged attention mechanism, which divides the query tensor into multiple pages, each processed independently. This operation helps to reduce computational redundancy and improve the efficiency of parallel computation. At the same time, the key and value tensor is expanded according to the grouping strategy to form key\_groups and value\_groups, so that each subgroup can obtain the corresponding key and value. And when dealing with very long sequences, it will first divide the
input data into multiple pages (partition) through the paged attention mechanism, and each page will only deal with a part of the sequence, as in (1), to deal with shorter sequences, i.e., it will slice the sequences into multiple pages according to the temporal position. Attention is computed separately for each block; longer sequences are processed as in (2), i.e., the output of each block is cached and then used in the computation of the next block. At this point indicates that the cached values from previous blocks are used when processing the current block. In using the group query mechanism, the computational performance and memory usage in large-scale sequence processing is optimized to ensure scalability in efficiently processing long sequences.
\par

\begin{equation}
\begin{aligned}
	Q_{\text{block}} &= X_{\text{block}} W_Q, \\
	K_{\text{block}} &= X_{\text{block}} W_K, \\
	V_{\text{block}} &= X_{\text{block}} W_V.
\end{aligned}
\end{equation}

where $X_{block}$ is the input tensor for the current statement block,and $W_{Q}$,$W_{K}$,and,$W_{V}$ are the weight matrices for the Query,Key, and Value computations, respectively.

\begin{equation}
	\begin{aligned}
		V_{cached}=concatenate(V_{block}^{(i)}+V_{block}^{(i-1)})
	\end{aligned}
\end{equation}

where $V_{block}^{(i)}$ represents the cached values from the current block, and $V_{block}^{(i-1)}$ represents the cached values from the previous block.

\par
Next calculates the attention weights and we use matrix multiplication. For each set of query and key combinations, the dot product is first computed and scaling is adjusted according to the scaling factor (scale). At this point, efficient tensor operations (e.g., torch.einsum) are used to speed up the matrix computation. For each computation between query heads (num\_heads) and key heads (num\_kv\_heads), the attention weights are further adjusted to avoid unnecessary autoregressive dependencies or attention computations at invalid locations by adding a mask (attn\_mask). Ultimately, these attentional weights will be used as the basis for subsequent operations. The following is the formula for calculating the attention weights, as in (3).

{\small
	\begin{equation}
		\begin{aligned}
	Attention\_Raw(i,\ j)=query(i,:)\times{key(j,:)}^\intercal+bias(i,j)
		\end{aligned}
	\end{equation}
}

where $query(i,:)$ represents the i-th row of the query matrix, and transpose operation ${key(j,:)}^\intercal$ enables standard vector multiplication. Finally, a bias term $bias(i,j)$ is added to the result, which is typically used to adjust the weights or offsets between different query-key pairs.
\par

The computed attention weights are normalized using Softmax, as in (4), to convert them into a probability distribution. The role of Softmax is to transform the similarity between each query and all keys into a weighted sum, where the weight of each key represents its importance to the query. This normalization step ensures that the sum of all weights under each query equals 1, forming a valid attention distribution.

\begin{equation}
	\begin{aligned}
		Softmax(x_i)=\frac{e^{x_i}}{\sum_{j}e^{x_j}}
	\end{aligned}
\end{equation}

where $x_i$ is the input value for the ii-th element.$e^{x_i}$ is the exponential of the input value for the ii-th element. $\sum_{j}e^{x_j}$ is the sum of the exponentials of all input values.
\par

Finally, based on the normalized attention weights, these weights are used to compute the weighted sum of the value tensor, as in (5). The output of each query is composed of the weighted combination of its corresponding keys and values. In this way, each query can retrieve the most relevant information from multiple keys and values, ultimately generating the output for that query. For grouped queries, this process independently performs the weighted sum for each subgroup of queries and key-value pairs, and then combines the results into the final output.

\begin{equation}
	\begin{aligned}
	{Attention}_{weight}=Softmax(\frac{{\rm QK}^\intercal+Mask}{\sqrt{{\rm head}_{size}}})
	\end{aligned}
\end{equation}

where ${\rm QK}^\intercal$ is dot product of the query matrix and the transpose of the key matrix. Mask is masking matrix to prevent certain elements from contributing to the attention scores.$\sqrt{{\rm head}_{size}}$ scaling factor to normalize the dot product

Algorithm 1 summarizes the Opt-GQA forward pass, including tensor reshaping, cache handling, GQA alignment, bias setup, and attention computation.

\begin{algorithm}[htbp]
	\caption{Opt-GQA Forward Pass}
	\begin{algorithmic}[1]
		\State \textbf{Input:} $Q$, $K$, $V$, $C_{K}$, $C_{V}$, $M$
		\par
		\State \textbf{Output:} $Attention_output$
		
		Reshape Q, K, V to dimensions [batch\_size, seq\_len, num\_heads, head\_size],
		[batch\_size, seq\_len, num\_kv\_heads, head\_size], [batch\_size, seq\_len, num\_kv\_heads, head\_size]
		\If{Caches provided}
			\State Reshape and store K, V into $C_{K}$, $C_{V}$
		\EndIf
		\If{$M.\text{is\_prompt}$}
			\If{Caches are empty}
				\If{$\text{num\_kv\_heads} \neq \text{num\_heads}$}
					\State Adjust $Q$, $K$, $V$ for GQA grouping
				\EndIf
		
				\If{M.attn\_bias is empty}
					\If{ALiBi slopes empty}
						\State Create causal and local attention masks
					\Else
						\State Create ALiBi bias
					\EndIf
					\State Assign bias to $M.\text{attn\_bias}$
				\EndIf
			\EndIf
		\EndIf
		\If{Using reference attention}
			\State Compute reference masked attention
		\Else
			\State Compute attention scores, apply bias and softmax
			\State Compute weighted sum for Output
		\EndIf
		\State Reshape Output to [batch\_size, seq\_len, hidden\_size]
		\State \Return{Output}
	 \end{algorithmic}
\end{algorithm}

\subsection{Optimization Strategies and Implementation Details}

To further enhance the performance of Opt-GQA on DCUs, we have adopted several optimization strategies, primarily covering memory allocation and management optimization, DCU kernel optimization, cache sharing and reuse, as well as load balancing and resource scheduling.

\begin{itemize}
	\item \textbf{Memory Allocation and Management Optimization}
	\par
	Pre-allocate fixed-size memory pools and organize key-value vectors in contiguous blocks to minimize fragmentation and improve access speed.
	
	\item \textbf{DCU Kernel Optimization}
	\par
	Leverage SIMD-based vectorized operations and optimize memory access patterns to reduce latency and maximize throughput.
	
	\item \textbf{Cache Sharing and Reuse}
	\par
	Enable intelligent cache reuse and consistency control during concurrent access, reducing redundant computation and prioritizing hot key-value pairs.
	
	\item \textbf{Load Balancing and Resource Scheduling}
	\par
	Employ dynamic scheduling based on real-time load and query distribution to ensure balanced resource utilization and system stability under high concurrency.

\end{itemize}

Algorithm 2 summarizing the main steps of memory management, DCU computation optimization, cache sharing, and resource scheduling.

\begin{algorithm}[htbp]
	\caption{Opt-GQA Optimization on DCU}
     \begin{algorithmic}
	\State \textbf{Input:} $Q$, $K$, $V$, $C_{K}$, $C_{V}$, $M$
	\State \textbf{Output:} $Attention\_output$
	
	\State Pre-allocate GPU memory pools to minimize allocation overhead
	\State Reshape Q, K, V for efficient processing
	
	\If{Caches are provided}
	\State Store reshaped K, V into $C_{K}$, $C_{V}$
	\EndIf
	
	\If{$M.\text{is\_prompt}$}
	\If{Caches are empty}
	\State Adjust Q, K, V for GQA grouping
	\State Assign attention bias masks
	\EndIf
	
	\If{Using reference attention}
	\State Compute reference masked attention
	\State \Return{return reshaped output}
	\Else
	\State Compute attention scores, apply bias and softmax
	\State Compute weighted sum for Output
	\EndIf
	\Else
	\State Compute context attention with caches
	\EndIf
	
	\State Reshape Output to [batch\_size, seq\_len, hidden\_size]
	\State \Return{Attention\_output}
\end{algorithmic}
\end{algorithm}

\section{EXPERIMENT AND RESULT ANALYSIS}
\subsection{Experimental Environment}
This study establishes a performance evaluation baseline based on the unoptimized vLLM \cite{b15} serving system, which serves as a reference point for analyzing the effectiveness of the proposed optimization strategies. To ensure the reliability and reproducibility of the experimental results, all experiments are conducted within a consistent and controlled hardware and software environment. The experiments are conducted on the HYGON DCU Z100 platform, equipped with 3840 compute cores for parallel thread execution, 32GB of HBM2 memory for efficient storage of intermediate activations and attention tensors. A series of quantized language models are employed in the evaluation, including LLaMa3-8B-GPTQ, LLaMa2-13B-GPTQ \cite{b16}, LLaMa-7B-GPTQ, LLaMa-13B-GPTQ, and LLaMa-Pro-8B-GPTQ \cite{b17}. The effectiveness of the optimization strategies is systematically assessed through a comprehensive analysis of key performance metrics, including latency, generation throughput, and all throughput.
\subsection{Experimental Results}
In this section, we present the experimental results of
the Opt-GQA mechanism implemented on DCUs. By comparing
it with traditional MHA, the experiments evaluate
improvements in computational efficiency, memory usage, and
model performance. We will showcase key data and analyses
that demonstrate the effectiveness and advantages of Opt-
GQA.
	
	\begin{figure}[htbp]
		\centerline{\includegraphics[width=\columnwidth]{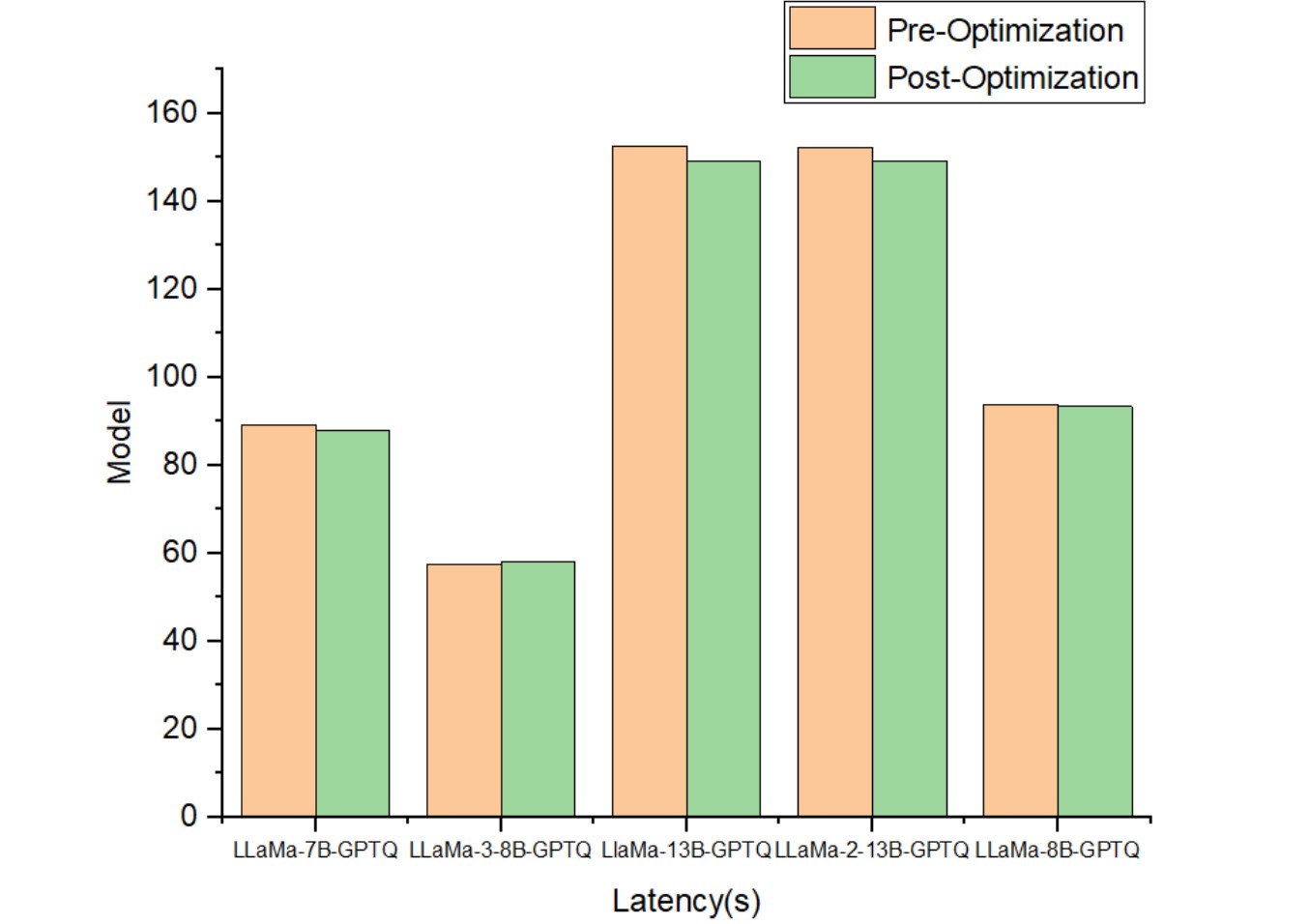}}
		\caption{Latency Impact of Opt-GQA Optimization}
		\label{fig2}
	\end{figure}
	
As shown in Fig. 2, after applying the Opt-GQA optimization, the inference latency of LLaMa-7B-GPTQ, LLaMa-3-8B-GPTQ, LLaMa-13B-GPTQ, LLaMa-2-13B-GPTQ, and LLaMa-8B-GPTQ saw improvements in inference latency of 1.33\%, -0.64\%, 2.35\%, 2.04\%, and 0.45\%, respectively. These results demonstrate that Opt-vLLM achieves varying degrees of latency optimization across different LLaMa model variants.\par It is worth noting that the latency of the LLaMa-3-8B-GPTQ model slightly increased, likely due to its architecture or suboptimal optimization. Tuning kernel parameters, caching, or thread granularity may reduce inference time.

	\begin{figure}[htbp]
		\centerline{\includegraphics[width=\columnwidth]{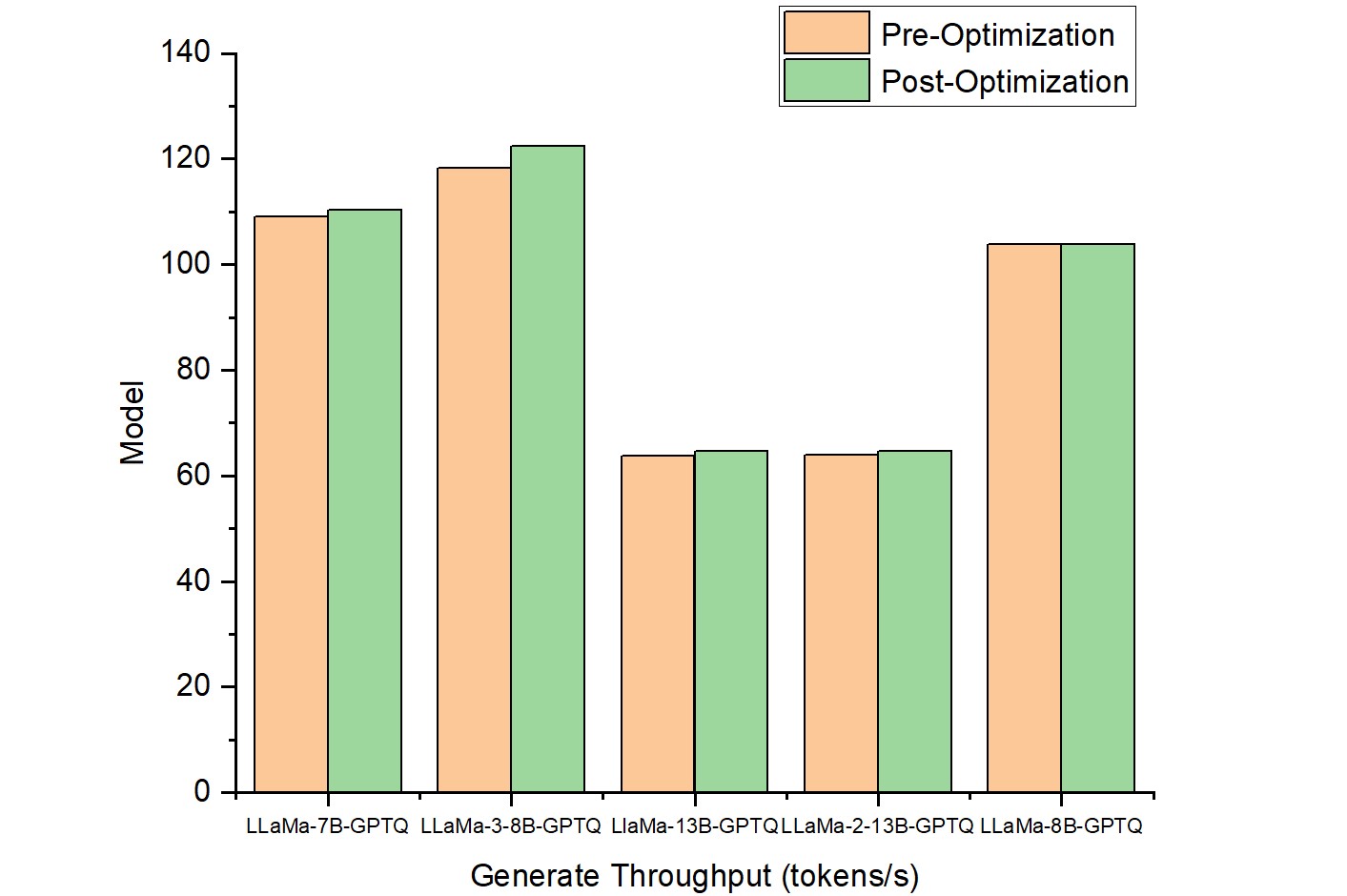}}
		\caption{Generation Throughput Impact of Opt-GQA Optimization}
		\label{fig3}
	\end{figure}
	
	As shown in Fig. 3, after applying the optimization, the generation throughput of LLaMa-7B-GPTQ, LLaMa-3-8B-GPTQ, LLaMa-13B-GPTQ, LLaMa-2-13B-GPTQ, and LLaMa-8B-GPTQ saw improvements in generation throughput of 1.17\%, 3.47\%, 1.72\%, 1.40\%, and 0.11\%, respectively. Although the improvement margins are limited, these results indicate that the optimization scheme can effectively enhance token processing efficiency without altering the model structure, demonstrating good generalizability and scalability.

	\begin{figure}[htbp]
		\centerline{\includegraphics[width=\columnwidth]{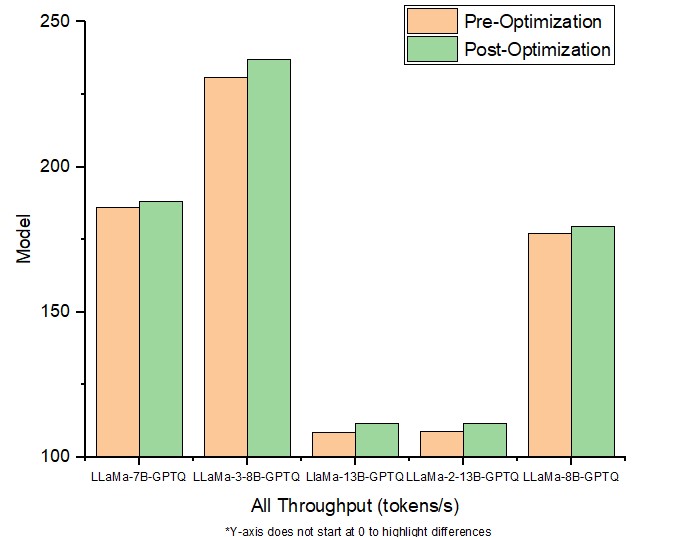}}
		\caption{All Throughput Impact of Opt-GQA Optimization}
		\label{fig4}
	\end{figure}
	
	As shown in Fig. 4, after applying the optimization, the all throughput of LLaMa-7B-GPTQ, LLaMa-3-8B-GPTQ, LLaMa-13B-GPTQ, LLaMa-2-13B-GPTQ, and LLaMa-8B-GPTQ saw improvements in all throughput of 1.07\%, 2.77\%, 2.70\%,2.29\%, and 1.46\%, respectively. Although the improvement margins are limited, these results indicate that the optimization scheme can effectively enhance token processing efficiency without altering the model structure, demonstrating good generalizability and scalability.

\subsection{Discussion}

Although Opt-GPTQ is generally stable and the grouped attention mechanism has already brought about noticeable performance improvements, there are still some details that can be further optimized. First, while latency has slightly decreased, the change is minimal, indicating room for further optimization, especially in certain scenarios where observed increases in latency suggest the need to refine the paging strategy. For instance, appropriately adjusting the group size, optimizing memory usage strategies, or even adopting alternative efficient attention mechanisms could help mitigate latency issues and increase throughput. Secondly, the slight decrease in generation throughput suggests that the system may face computational resource bottlenecks or uneven scheduling when handling generation tasks. As models scale up, potential bottlenecks in memory and computational resources may limit further performance gains. To enhance performance, optimizing generation throughput is an important direction. Considerations could include parallel computation, resource allocation optimization, or improvements at the algorithmic level to increase generation throughput while maintaining system stability.

\section{CONCLUSION}
Opt-GQA mechanism enhances the efficiency of large-scale language models by minimizing computational redundancy and memory usage through query grouping and the use of shared key-value vectors. Its integration into vLLM models on DCUs demonstrates improved resource utilization, scalability, and parallel computing capabilities. Experimental results reveal significant gains in throughput and memory efficiency, establishing this method as a reliable solution for inference tasks. Although minor latency challenges remain, Opt-GQA provides a strong foundation for future optimizations to address increasingly complex AI workloads.

\section*{Acknowledgment}

This work was support in part by the China NSFC under Grant 62072287; in part by the China NSFC under Grant W2412090; in part by the ghFund under Grant 202407027775; in part by the project ZR2024ME230 supported by Shandong Provincial Natural Science Foundation.

\end{document}